\documentclass[11pt,a4paper]{article}
\usepackage{jcappub,epsfig,multicol,bbm,latexsym,graphicx,subfigure}

\def\aap{Astron.\ Astrophys.\ }
\def\apj{Astrophys.\ J.\ }
\def\apjl{Astrophys.\ J.\ Lett.\ }

\def\mnras{Mon.\ Not.\ Roy.\ Astron.\ Soc.\ }

\def\prc{Phys.\ Rev.\ C\ }
\def\prd{Phys.\ Rev.\ D\ }
\def\prl{Phys.\ Rev.\ Lett.\ }

\def\jcap{J.\ Cosmol.\ Astropart.\ Phys.\ }

\title{Secondary cosmic-ray nucleus spectra disfavor particle 
transport in the Galaxy without reacceleration}

\author{Qiang Yuan$^{a,b,c}$\footnote{Corresponding author.}, 
Cheng-Rui Zhu$^{a,d}$,
Xiao-Jun Bi$^{e,d}$\footnote{Corresponding author.},
Da-Ming Wei$^{a,b}$}

\affiliation{
$^a$Key Laboratory of Dark Matter and Space Astronomy, Purple Mountain
Observatory, Chinese Academy of Sciences, Nanjing 210008, China\\
$^b$School of Astronomy and Space Science, University of Science and
Technology of China, Hefei, Anhui 230026, China\\
$^c$Center for High Energy Physics, Peking University, Beijing 100871, China\\
$^d$University of Chinese Academy of Sciences, Beijing 100049, China\\
$^e$Key Laboratory of Particle Astrophysics, Institute of High Energy
Physics, Chinese Academy of Sciences, Beijing 100049, China}
\emailAdd{yuanq@pmo.ac.cn}
\emailAdd{zhucr@pmo.ac.cn}
\emailAdd{bixj@ihep.ac.cn}
\emailAdd{dmwei@pmo.ac.cn}

\abstract{
The precise observations of Galactic cosmic ray fluxes of the secondary 
family, such as Li, Be, B, are expected to have significant implications
on our understanding of the cosmic ray origin and propagation. Here we
employ the recent very precise measurements of those species by the Alpha 
Magnetic Spectrometer on the International Space Station, together with
their parent species (C and O), as well as the data collected by the
Voyager-1 spacecraft outside the heliosphere and the Advanced Composition
Explorer, to investigate the propagation of cosmic rays in the Milky Way.
We consider the diffusion of cosmic rays plus reacceleration or
convection effect during the propagation, and find that the reacceleration 
model can fit the data significantly better than the convection model.
We further find that for the reacceleration model, the spectral hardenings 
of both the primary and secondary particles can be well described by the 
injection hardening without including additional propagation hardening.
This is due to that the reacceleration effect results in a steeper 
secondary-to-primary ratio at low energies, and can thus naturally 
reproduce the fact that the secondary spectra harden more than the 
primary spectra found by AMS-02.}

\arxivnumber{1810.03141}

\begin{document}
\maketitle
\flushbottom

\section{Introduction}

It has been well established that charged cosmic rays (CRs) propagate
diffusively in the Milky Way, and interact with the interstellar medium,
fields, and plasma waves. Such interactions would leave imprints on
their spectra and produce secondary particles and radiation
\cite{2007ARNPS..57..285S}. Precise measurements of CR energy spectra,
particularly the secondary class such as Li, Be, B or Sc, Ti, V, Cr, 
Mn are crucial to probing the propagation and interactions of CRs.
Using improved data acquired in recent years by, e.g., PAMELA and 
AMS-02, we do have significantly improved constraints
on the CR propagation parameters \cite{2011ApJ...729..106T,
2015JCAP...09..049J,2016ApJ...824...16J,2016PhRvD..94l3019K,
2016PhRvD..94l3007F,2017PhRvD..95h3007Y,2018PhRvD..97b3015N,
2018JCAP...01..055R,2019SCPMA..6249511Y,2019PhLB..789..292W,
2020PhRvD.101b3013E,2020arXiv200211406W}. 
Nevertheless, the big patterns of the CR propagation are still under debate, 
and those constraints on the model parameters are model-dependent.

Other than the diffusion in turbulent magnetic fields and the inelastic 
collisions with the interstellar gas, CR particles are widely postulated
to experience convective propagation \cite{1976ApJ...208..900J} which is 
believed to be the source driving Galactic stellar winds and/or stochastic 
reacceleration by randomly moving magnetohydrodynamic (MHD) waves 
\cite{1994ApJ...431..705S}. Previous measurements are not precise 
enough to discriminate these two classes of model configurations, 
although several recent studies showed hints that the reacceleration
model was somehow favored compared with the convection one
\cite{2017PhRvD..95h3007Y,2019SCPMA..6249511Y}. There were also
uncertainties from the entanglement of the propagation model with 
assumptions of the source injection and solar modulation.

Here we investigate this problem with the most recently published 
data on Li, Be, B and their parent nuclei C and O by AMS-02 
\cite{2017PhRvL.119y1101A,2018PhRvL.120b1101A}. To better constrain 
the model, the low-energy measurements of these nuclei by Voyager-1 
out of the heliosphere \cite{2016ApJ...831...18C}, and the medium-energy 
measurements by the ACE-CRIS instrument at top of the atmosphere are 
also used. We keep our studies in a relatively simple framework
of particle transportation with uniform, isotropic diffusion of CRs,
although spatial variations and/or anisotropic diffusion may be
indicated by some recent observations \cite{2012ApJ...752L..13T,
2016ApJ...819...54G,2017JCAP...10..019C}. As for the solar modulation
effect of CRs in the heliosphere, we employ the force-field approximation
\cite{1968ApJ...154.1011G}. In spite that more complicated model settings
may be relevant when confronting all available data such as the diffuse 
$\gamma$-rays, our results show that this simple framework may be enough 
to account for the CR data which are largely observed near the Earth.
The function form of the injection spectrum is poorly known. Typically 
broken power-law forms are assumed to describe the injection spectrum of 
accelerated CRs. In this work we adopt a non-parametric, spline 
interpolation method to describe the particle's injection spectrum.
This can also minimize the impact of an improper assumption of the
injection spectrum on the determination of the propagation parameters. 

\section{Methodology}

The propagation of CRs is described by a set of coupled diffusion
equations, including the energy losses and fragmentations due to
interactions in the medium, the reacceleration and/or the convection 
effects. For details about the propagation equations one can refer
to Ref.~\cite{2007ARNPS..57..285S}. With proper simplifications the
propagation equations can be solved analytically (e.g., 
\cite{2001ApJ...555..585M}).
However, for general purposes and realistic astrophysical ingredients, 
a numerical way is usually necessary \cite{1998ApJ...509..212S,
1998ApJ...493..694M,2008JCAP...10..018E,2017JCAP...02..015E}. 
We use the numerical code {\tt GALPROP} to calculate the propagation 
of CRs \cite{1998ApJ...509..212S,1998ApJ...493..694M}. A cylindrically
symmetric geometry of the Galaxy is assumed, with a maximum radius
$r_{\rm max}=20$ kpc and a height of $2z_h$ with $z_h$ free to be
fitted. It has been shown that a three-dimensional configuration of
the propagation geometry does have some impacts on the results of
propagation \cite{2013PhRvL.111b1102G,2017ApJ...846...67P,
2018ApJ...856...45J}. We expect that such effects will affect
simultaneously on both models we are going to compare with, and thus
neglect such a complexity.

The diffusion coefficient is parameterized as $D(\rho)=\beta^{\eta}D_0
(\rho/\rho_0)^{\delta}$, where $\rho$ is the rigidity of CR particles,
$D_0$ is (approximately) the diffusion coefficient at $\rho_0=4$ GV,
$\delta$ is the rigidity-dependence slope, $\beta$ is the velocity in 
unit of light speed, and the $\beta^{\eta}$ term is employed to
empirically describe possible resonant scatterings of CRs off the
MHD waves \cite{2006ApJ...642..902P,2010APh....34..274D}.
We will also test the case with a high-rigidity break of the diffusion
coefficient, which was suggested to account for the spectral breaks
of both the primary and secondary nuclei around 200 GV, e.g., 
\cite{2017PhRvL.119x1101G}. In this case two more parameters of the
diffusion coefficient, $\rho_{\rm br}$ and $\delta'$, are added.
The reacceleration effect is described by a diffusion in the momentum
space, and the Alfven velocity ($v_A$) of the MHD waves is employed to 
characterize the strength of reacceleration \cite{1994ApJ...431..705S}.
The convection is assumed to be perpendicular to the Galactic plane,
and the convection velocity is parameterized as a linear function of
the vertical height to the Galatic plane, $V_c=z\cdot dV_c/dz$.

To minimize the impact of the parameterization of the injection 
spectrum of primary CRs, we adopt a non-parametrized method by means 
of spline interpolation among a few chosen rigidity knots
\cite{2016A&A...591A..94G,2018ApJ...863..119Z}. The (logarithmical) 
flux normalizations at such knots are fitted together with other free 
parameters. In this work we take 7 rigidity knots\footnote{We also test 
the fits with different numbers of knots, which result in slightly 
different values of the fitting model parameters and $\chi^2$, without 
affecting all the conclusions in this work.} logarithmically evenly 
distributed between 0.2 GV and 2 TV, which covers the energy ranges
of observations from Voyager-1 to AMS-02. The abundance ratio between 
primary nuclei C and O is described by a constant factor $\xi_{\rm O}$. 
In addition, there might be uncertainties of fragmentation cross 
sections to produce secondary nuclei \cite{2018PhRvC..98c4611G}, 
which would degenerate with the propagation parameters. In this work 
we fix the cross section of the Boron production, and multiply two 
constants\footnote{The uncertainties of the cross sections may be
energy-dependent (e.g., \cite{2019A&A...627A.158D}). Here we neglect such 
complexitites due to the lack of precise knowledge about the cross sections.} 
$\xi_{\rm Li}$ and $\xi_{\rm Be}$ to the predicted fluxes of Li and Be. 
Finally, the spatial distribution of CRs is assumed to follow the 
observed distribution of supernova remnants \cite{2011ApJ...729..106T}. 

To link the ACE-CRIS and AMS-02 measurements near the Earth with
the Voyager-1 measurements outside the heliosphere, the force-field
solar modulation model is employed. The modulation potential is set
as a free parameter. Note that we extracted the ACE-CRIS data with 
the same time period as that by the AMS-02 measurements of nuclei
Li, Be, B, C, O, i.e., from May, 2011 to May, 2016
\cite{2018ApJ...863..119Z}.

In total there are 16 or 18 free parameters in the model: 5 or 7 for the 
modeling of propagation ($D_0$, $\delta$, $\eta$, $z_h$, $v_A$ or $dV_c/dz$),
including additionally ($\rho_{\rm br}$, $\delta'$) for the case with a
break of the diffusion coefficient, 7 for the injection spectrum, 3 for 
normalizations of Li, Be, and O, and 1 for solar modulation. 
It is challenging to fit all of those parameters together, due to the 
degeneracies among many of them. Therefore we adopt an iterative approach. 
We start with a given set of propagation parameters (e.g., those given in 
Ref.~\cite{2019SCPMA..6249511Y}), and fit the injection parameters plus 
the normalization factors. Then we take the best-fit injection and 
normalization parameters as inputs, and re-fit the propagation parameters. 
We repeat the procedure until the results get stable. We use the 
{\tt CosRayMC} tool which employs the Markov Chain Monte Carlo method 
to do the fits \cite{2010PhRvD..81b3516L,2012PhRvD..85d3507L}. 

\section{Results}

The data used in this work include the fluxes of Li, Be, B, C, and O measured 
by AMS-02 \cite{2017PhRvL.119y1101A,2018PhRvL.120b1101A} and Voyager-1 
\cite{2016ApJ...831...18C}, the B, C, and O fluxes measured by the ACE-CRIS 
\cite{2018ApJ...863..119Z}. The old measurements of the Beryllium
ratios \cite{1988SSRv...46..205S,1998ApJ...501L..59C,1999ICRC....3...41L,
2001ApJ...563..768Y,2004ApJ...611..892H} are also included to give a 
loose constraint on the halo height. We are aware that it is crucial to 
using the simultaneous measurements in this study. However, since the 
new measurements of the Beryllium ratios are not available yet, we can
only use those old data at the current stage. For those old data of 
$^{10}$Be/$^9$Be ratios, we apply a force-field solar modulation with a 
modulation potential of $\Phi\approx\Phi_{\rm AMS-02}-0.2=0.5$ GV 
\cite{2017PhRvD..95h3007Y}.

\begin{figure}[!htb]
\centering
\includegraphics[width=0.48\textwidth]{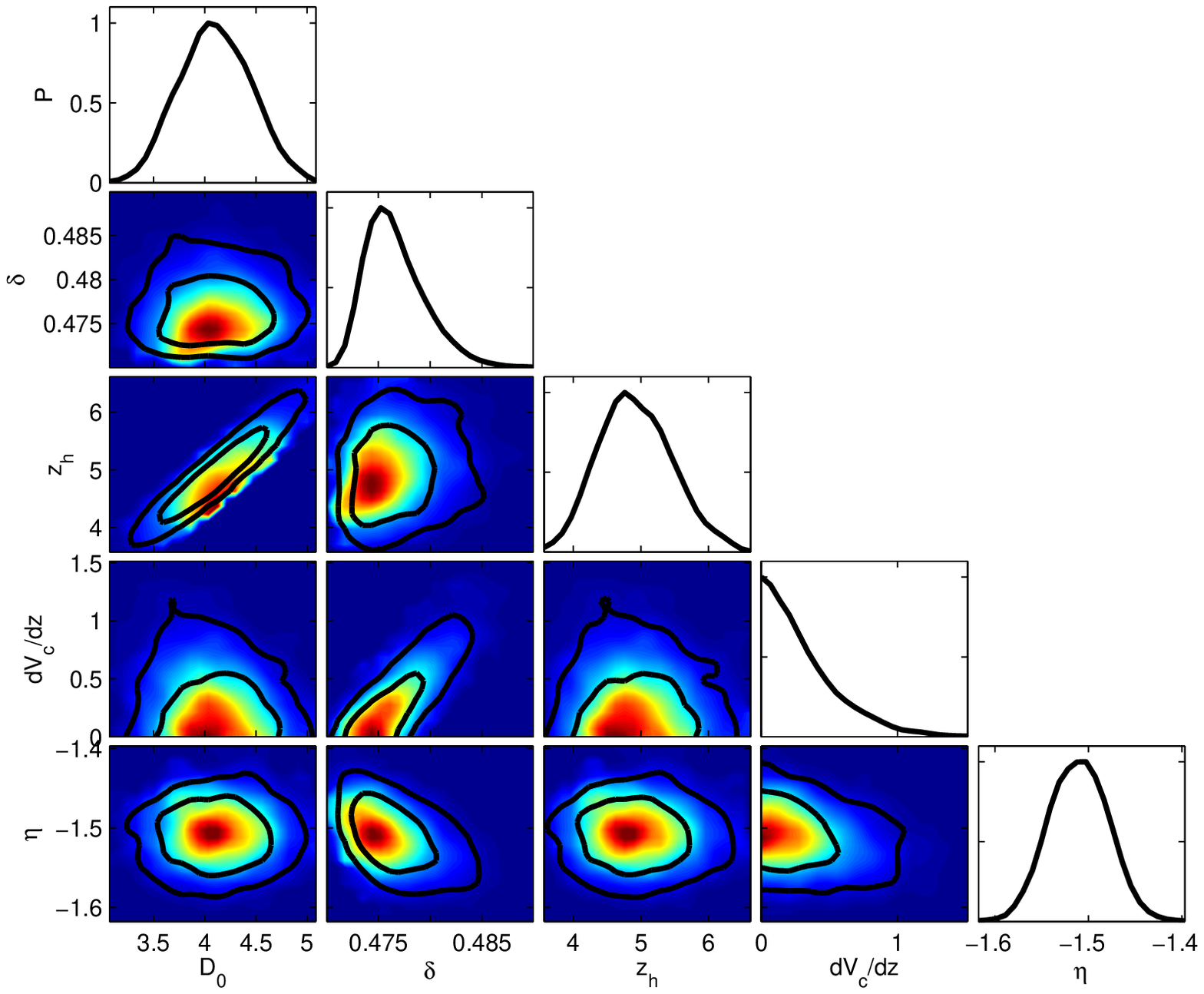}
\includegraphics[width=0.48\textwidth]{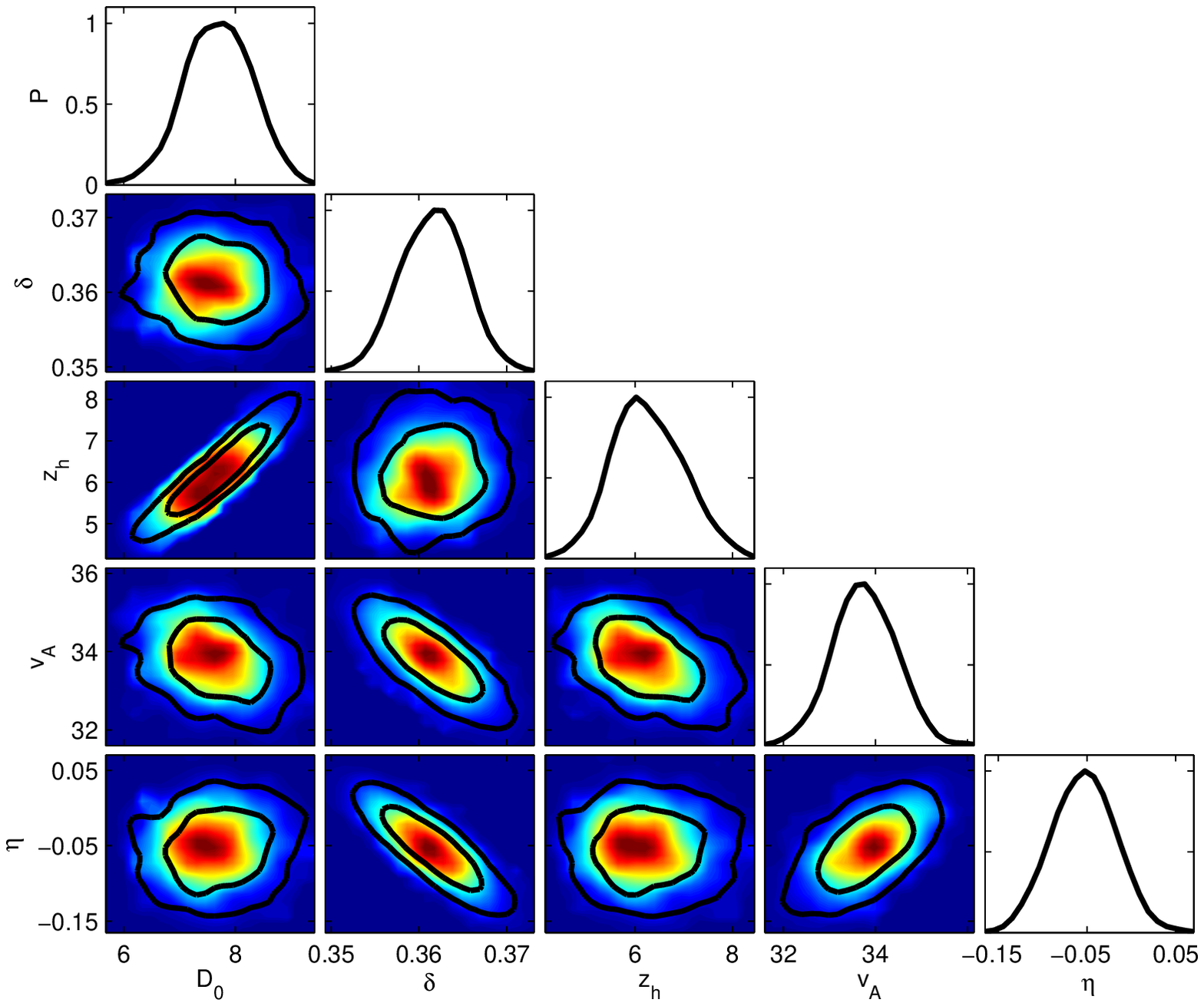}
\caption{The 1-D and 2-D distributions of the propagation parameters
for the convection (left) and reacceleration (right) models. The units
of the parameters are given in Table \ref{table:para}.
\label{fig:prop_tri}}
\end{figure}

The fitting results of the main propagation model parameters are given in 
Table \ref{table:para}. The one-dimensional (1-D) and two-dimensional (2-D)
distributions of the propagation parameters are shown in 
Fig.~\ref{fig:prop_tri}. It is clear that the reacceleration model fits 
the data significantly better than the convection model. For the convection 
model, the minimum $\chi^2$ value is about $445.9$ for a number of 
degree-of-freedom (dof) of 403, which gives a $p$-value of $0.07$. 
As for the comparison between the reacceleration and convection models, 
the Akaike information criterion (AIC) gives a difference of the AIC 
values of 140.5, which suggests that the convection model is about 
$\exp(-140.5/2)=3.1\times10^{-31}$ times as probable as the 
reacceleration one. We note that the use of the $\chi^2$ values to 
compare the models or judge the goodness-of-fit of specific models 
should be cautious, since the reported systematic uncertainties of the 
measurements have been added in quadrature with the statistical 
errors and the possible correlation among systematic uncertainties 
\cite{2019A&A...627A.158D,2020arXiv200504237H} have not been taken into 
account in the calculation of the $\chi^2$. The chi-squared value of 
the reacceleration model is 305.4/403, which seems to be a too good fit. 
We expect that this may also be related with the correlated systematical 
uncertainties. A full description of the covariance matrix is currently 
challenging without knowing the details of the analysis.

\begin{table}
\centering
\caption {Posterior mean and $68\%$ credible uncertainties of the main
model parameters.}
\begin{tabular}{cccccccc}
\hline
\hline
Parameter                    & Unit                           & Convection        & Reacceleration    \\
\hline
$D_0$                        & $(10^{28}{\rm cm^2s^{-1}})$    & $4.10 \pm 0.34$   & $7.69 \pm 0.60$   \\
$\delta$                     &                                & $0.477 \pm 0.003$ & $0.362 \pm 0.004$ \\
$z_h$                        & $({\rm kpc})$                  & $4.93 \pm 0.53$  & $6.27 \pm 0.72$   \\
$v_A$                        & $({\rm km\,s^{-1}})$           & ...               & $33.76 \pm 0.67$    \\
$dV_c/dz$                    & $({\rm km\,s^{-1}\,kpc^{-1}})$ & $<0.86^\dagger$   & ...               \\
$\eta$                       &                                & $-1.51 \pm 0.03$  & $-0.05 \pm 0.04$  \\
$\chi^2_{\rm min}/{\rm dof}$ &                                & $445.9/403$         & $305.4/403$       \\ \hline
$\xi_{\rm Li}$ &   & $1.18\pm0.01$  & $1.18\pm0.01$ \\
$\xi_{\rm Be}$ &   & $1.00\pm0.01$  & $1.00\pm0.01$ \\
$\Phi$ & (GV)  &   $0.742\pm0.010$  & $0.705\pm0.009$ \\ \hline\hline
\end{tabular}\\
$^\dagger$95\% upper limit.
\label{table:para}
\end{table}

\begin{figure*}[!htb]
\centering
\includegraphics[width=0.48\textwidth]{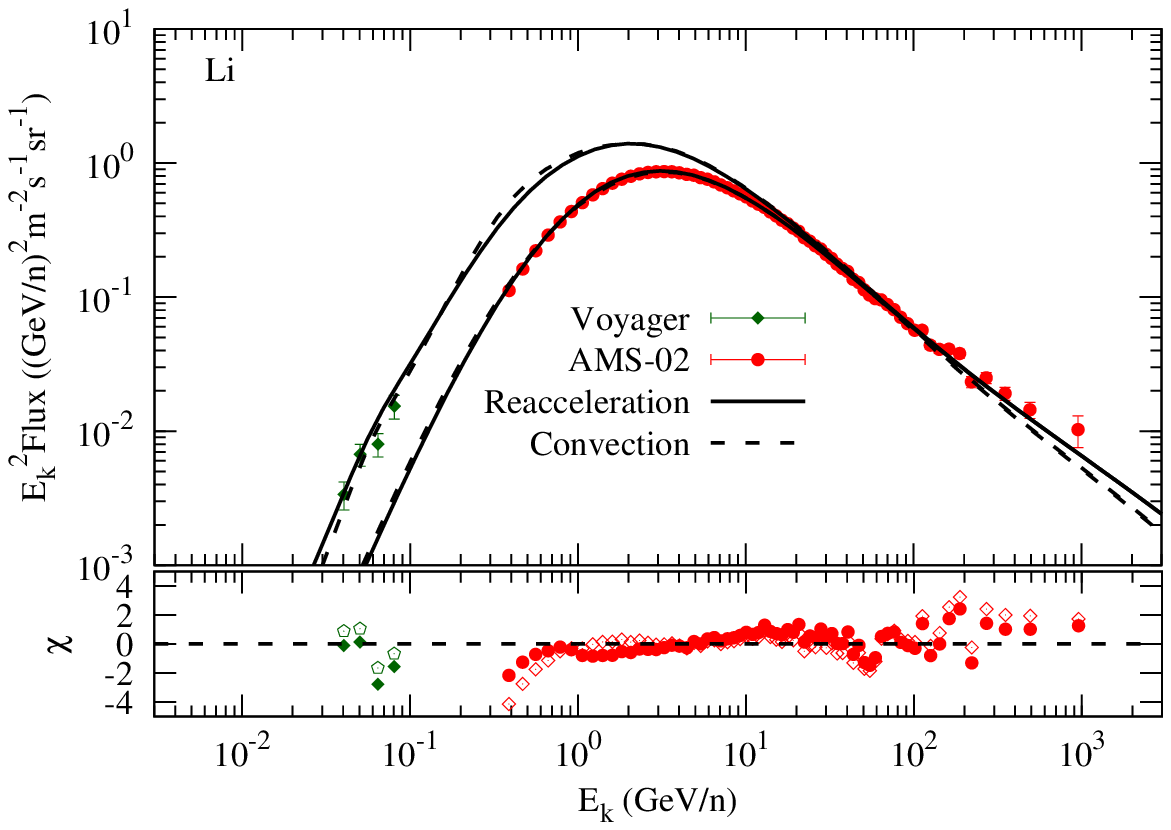}
\includegraphics[width=0.48\textwidth]{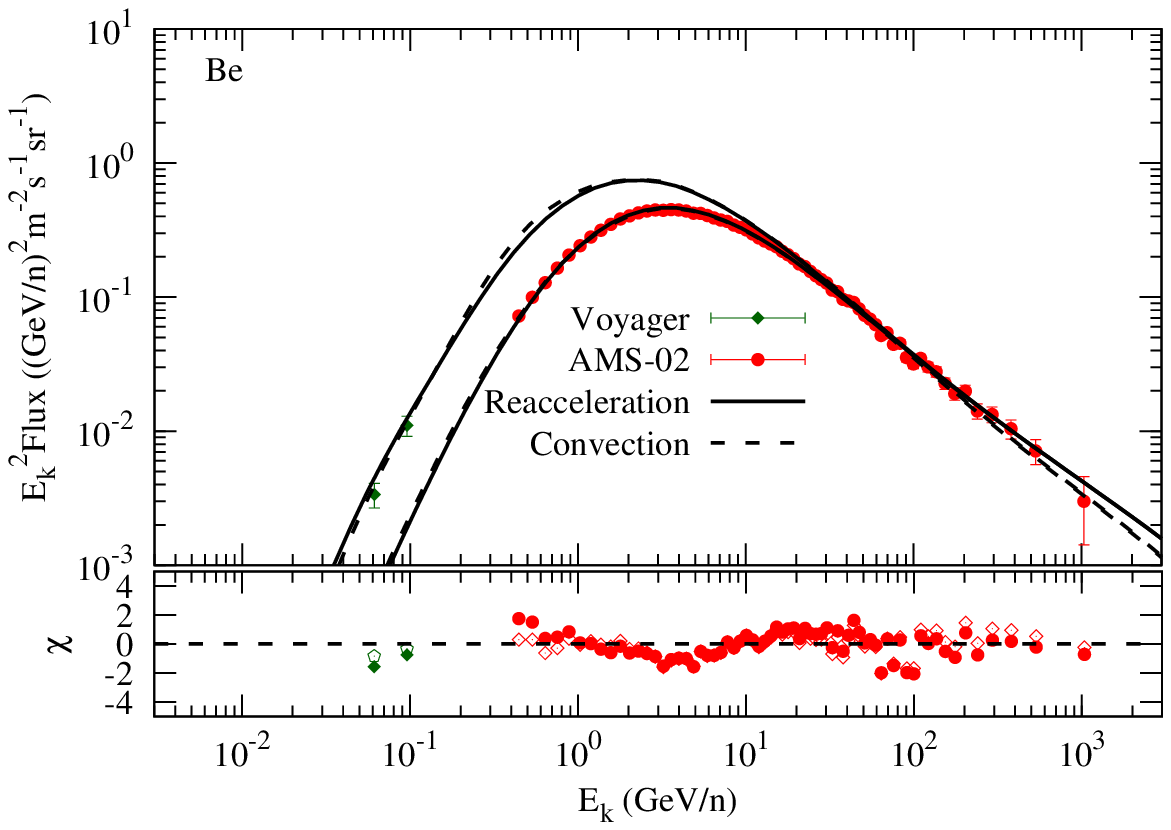}
\includegraphics[width=0.48\textwidth]{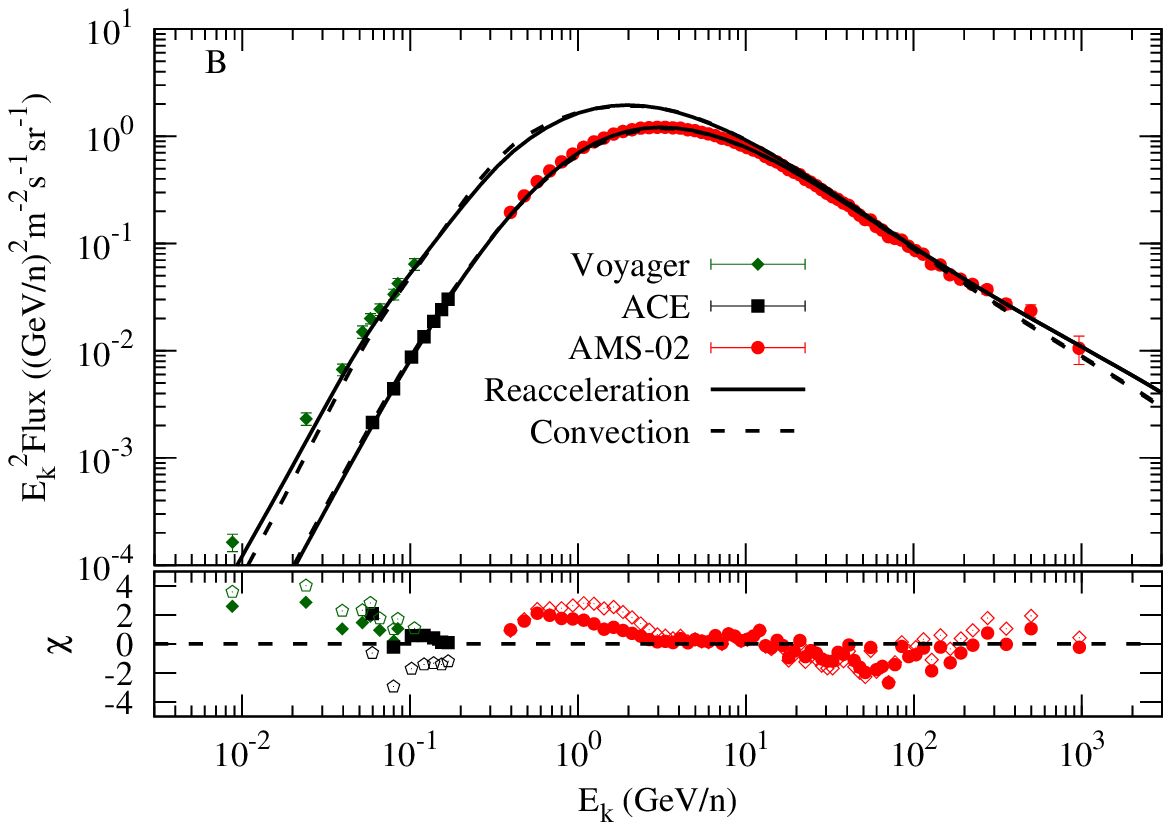}
\includegraphics[width=0.48\textwidth]{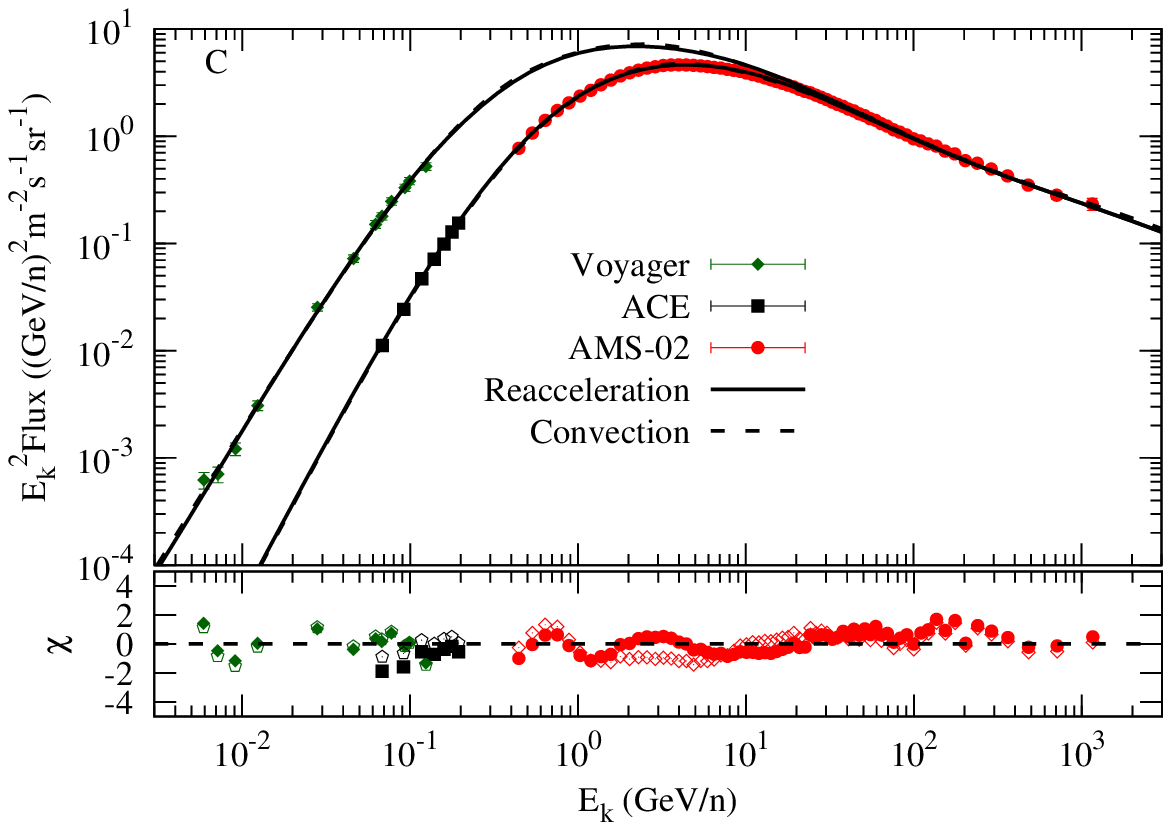}
\includegraphics[width=0.48\textwidth]{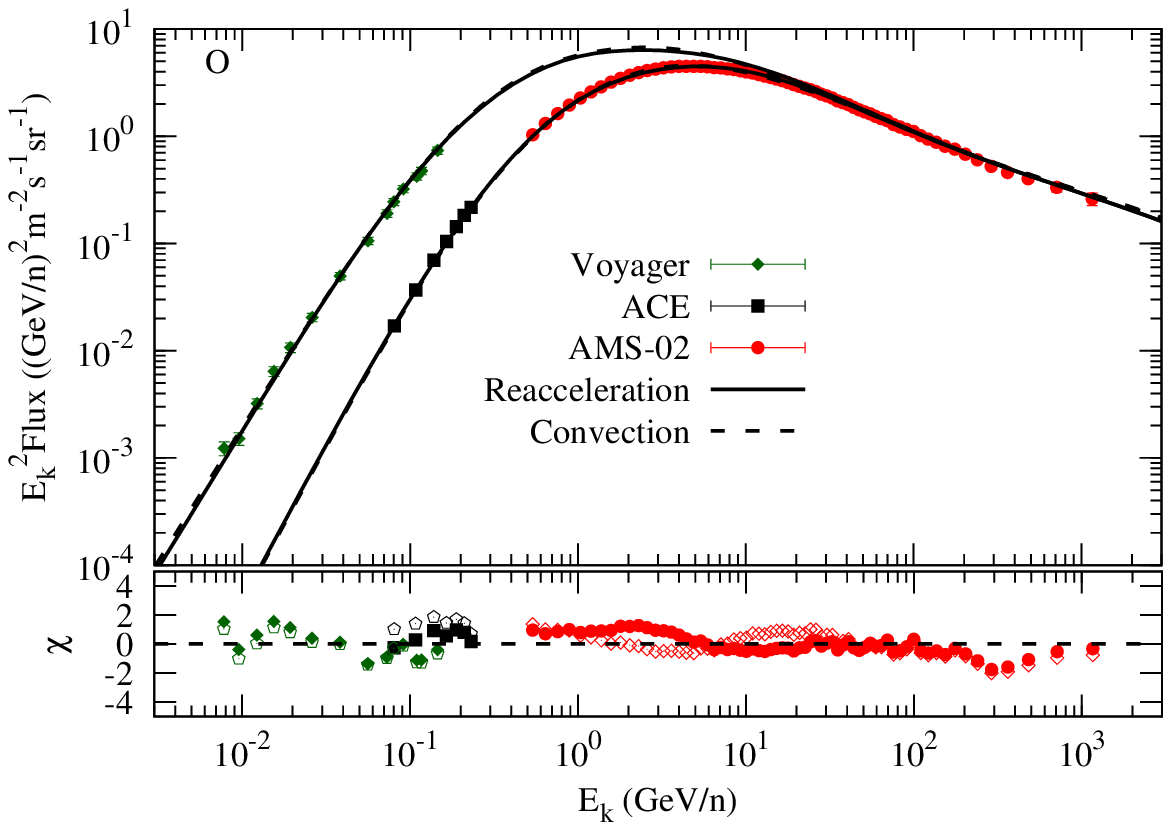}
\caption{Comparison between the best-fit model fluxes and the measurements. 
The lower sub-panel in each panel shows the residuals (filled for the 
reacceleration model and open for the convection model) defined as 
(data$-$model)/error. 
\label{fig:spec}}
\end{figure*}

Fig.~\ref{fig:spec} shows the comparison between the best-fit model
calculated fluxes and the data, for Li, Be, B, C, and O nuclei, 
respectively. We find that the convection model gives on 
average larger residuals than the reacceleration model. Furthermore, 
the value of $\delta$ is larger in the convection model than that in 
the reacceleration model. It is known that the reacceleration effect 
would make the bump feature of the secondary-to-primary ratio more 
prominent, and thus $\delta$ can be smaller \cite{1998ApJ...509..212S}. 
Therefore the predicted secondary fluxes by the convection model are 
generally softer than that by the reacceleration one.

\begin{figure}[!htb]
\centering
\includegraphics[width=0.6\textwidth]{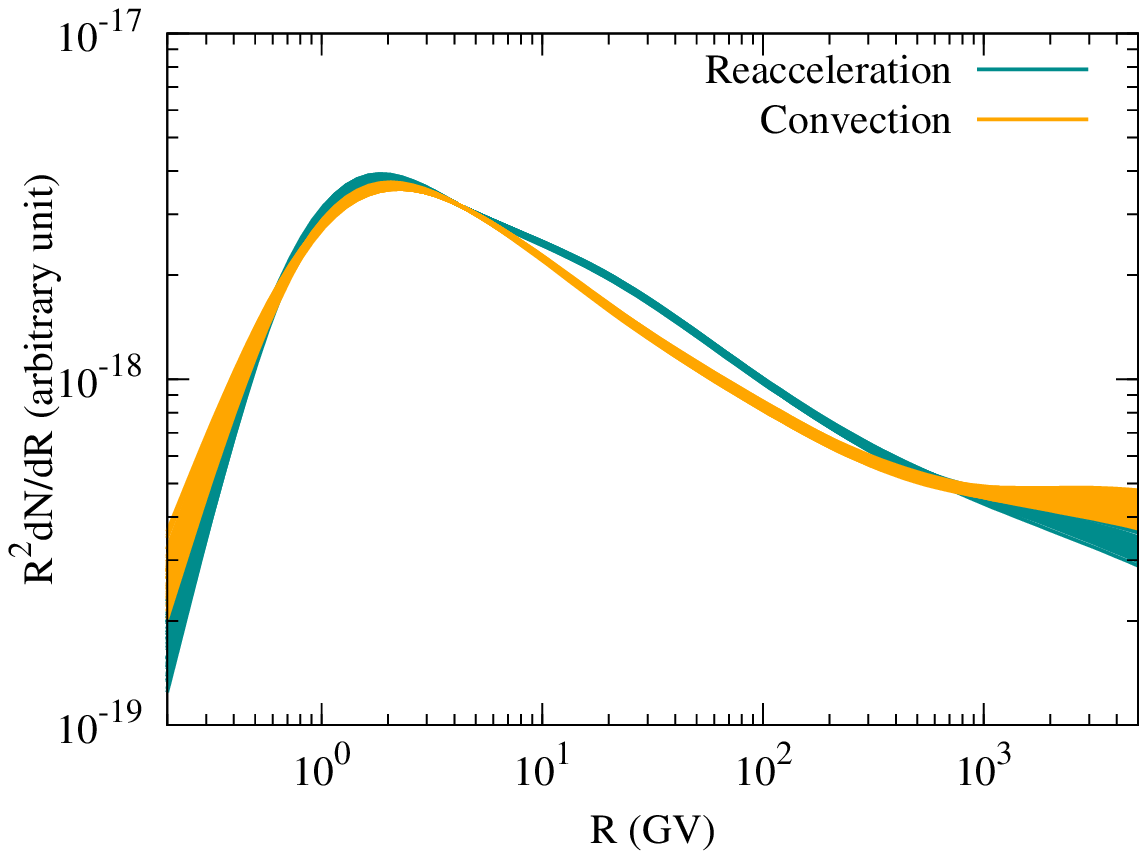}
\caption{Injection spectra for the convection and reacceleration models.
For each model we select 50 samples within the 95\% credible region of
the parameter space.
\label{fig:spec_inj}}
\end{figure}

The spectral hardenings above a few hundred GeV/n are found in both the 
primary and secondary fluxes \cite{2007BRASP..71..494P,2010ApJ...714L..89A,
2011Sci...332...69A,2015PhRvL.114q1103A,2015PhRvL.115u1101A,
2017PhRvL.119y1101A,2018PhRvL.120b1101A}.
A direct fitting to the secondary-to-primary ratios results in a change
of the slopes for rigidity intervals of $60.3-192$ and $192-3300$ GV
\cite{2018PhRvL.120b1101A}, suggesting that a propagation mechanism
is responsible to the spectral hardenings \cite{2012PhRvL.109f1101B,
2012ApJ...752L..13T,2016ApJ...819...54G}. Similar conclusion was also 
found in Ref.~\cite{2017PhRvL.119x1101G} for a fitting to the B/C data 
taking into account the propagation model.

In above we assume a single power-law form of the diffusion coefficient 
at high rigidities. Clear spectral hardenings can be seen in 
the fitting results for both primary and secondary nuclei (see 
Fig.~\ref{fig:spec}), which are expected to be due to the hardening
of the injection spectrum. Fig.~\ref{fig:spec_inj} shows the 95\% range
of the injection spectra for the convection and reacceleration models. 
Note that the injection spectra are normalized at $\sim4.3$ GV to show
the spectral shapes. These non-parametric injection spectra turn out 
to be similar with broken power-laws usually assumed. The spectra 
experience a softening at $\sim$GV rigidities and a hardening above 
several hundred GV. The physical origin of such spectral shapes would 
be very important in understanding the acceleration and/or confinement
of CRs at source. Note that the $\gamma$-ray emission from supernova
remnants also suggests broken power-law forms of particles around GeV
energies \cite{2013Sci...339..807A}. Possible physical mechanisms 
include the strong ion-neutral collisions near the shock fronts
\cite{2011NatCo...2E.194M} or the escape of particles from/into
finite-size regions \cite{2010MNRAS.409L..35L,2011MNRAS.410.1577O}.
The high-energy hardening may be due to the superposition of various
sources \cite{2011PhRvD..84d3002Y}, or the non-linear acceleration
\cite{2013ApJ...763...47P}.

We then check that whether the data require an additional break of
the diffusion coefficient or not. Two more parameters, the break
rigidity $\rho_{\rm br}$ and the high energy slope $\delta_{\rm he}$,
have been added in the model\footnote{Note that the momentum diffusion
coefficient depends on the parameter $\delta$ \cite{1994ApJ...431..705S}.
Here the break rigidity is restricted to be larger than 100 GV, where
the reacceleration effect is expected to be small. Therefore only the
low rigidity slope $\delta$ is relevant to the reacceleration.}. 
We find that for the reacceleration model, the addition of the break 
of the diffusion coefficient improves the fitting very slightly (with 
a minimum $\chi^2_{\rm br}$ value of $303.7$). For the convection model, 
however, a moderate improvement is found ($\chi^2_{\rm br}=389.6$).
Such a difference is anticipated as can be seen from Fig.~\ref{fig:spec}.
For the reacceleration model, the reacceleration effect makes the low 
energy secondary-to-primary ratios steeper than that at high energies,
which automatically gives the hardening behaviors as found by AMS-02.
Also the $\delta$ value of the rigidity-dependence of the diffusion
coefficient of the reacceleration model is relatively small, which 
is close to the high energy slope of the secondary-to-primary ratios.
On the other hand, a larger value of $\delta$ in the convection model
is required, basically to fit the low energy data. In this case an 
additional hardening of the diffusion coefficient at high energies 
is required. In this sense our results are consistent with that given 
in Ref.~\cite{2017PhRvL.119x1101G}, in which a minor reacceleration 
effect was assumed (with a prior range of $[0,10]$ km s$^{-1}$ for $v_A$) 
and a relatively larger value of $\delta\sim(0.5-0.7)$ was obtained. 
We regard this as a further support of the reacceleration model being
a natural explanation of the hardenings of the secondary-to-primary
ratios revealed by the AMS-02 measurements.

From the fittings we find that a re-normalization factor of $\sim1.2$
is required for the Li component. For Be nuclei the re-normalization
factor is very close to 1. This is possibly due to the uncertainties of 
the fragmentation cross sections of nuclei, either for the productions of 
Li or B. Note, however, as pointed out in Ref.~\cite{2020ApJ...889..167B}
it may be unlikely that the cross sections for all the channels to produce 
Li nuclei are lower by $\sim20\%$ at the same time. An alternative
explanation is the existence of primary acceleration sources of Li
which are presumably $^7$Li sources such as novae or type Ia
supernovae \cite{2018PhRvL.120d1103K,2020ApJ...889..167B}.

\section{Summary and discussion}

In this work we study the CR propagation models with the newest precise 
measurements of both the primary and secondary CR fluxes by AMS-02, 
ACE-CRIS, and Voyager-1. The framework as adopted in the GALPROP
tool has been employed, which inculdes the diffusion, energy losses,
and fragmentations of CRs. We focus on two different propagation
models, one with reacceleration of CRs due to scattering off randomly
moving MHD waves in the Milky Way and the other with global convective
propagation from the Galactic disk to the poles. To minimize the 
effect of the assumption of injection spectrum on the propagation
study, we use a non-parametric method with spline interpolation to 
describe the injection spectrum.

With those approaches, we find that the propagation model with
reacceleration fits the data significantly better than that with
convection. The best-fit $\chi^2$ value of the reacceleration model 
is smaller by about 140.5 than the convection model, with the same 
numbers of degrees of freedom. This conclusion can only be achieved 
with significantly improved measurements of the data in a very wide
energy range. Also the improvement of the analysis method with a 
global fitting tool and the elimination of dependence of the source 
injection models are helpful. Using the PAMELA data of $Z\leq2$ 
nuclei, Ref.~\cite{2019PhLB..789..292W} studied the CR propagation,
which also showed that the model with reacceleration fits the data
better than that of convection, although a combination of 
reacceleration and convection seems to give the best fit. 
In Refs.~\cite{2019A&A...627A.158D,2020arXiv200211406W}, the
authors used the AMS-02 data of secondary-to-primary ratios to test
the propagation models under the semi-analytical model, and found 
that the model with reacceleration and convection or the purely
diffusion one can describe fairly equally well of the data.
We note that there are some differences of the model settings and
data usage between their works and ours. For example, the propagation
geometry is different between two works (one-dimension versus two-dimension). 
The energy-dependent treatment of the uncertainties of the fragmentation 
cross sections in Refs.~\cite{2019A&A...627A.158D,2020arXiv200211406W}, 
and the usage of the low-energy data in this work may also lead to
differences. We expect that further efforts in clarifying effects of
those different approaches and reducing the uncertainties of the
cross sections will be particularly helpful in understanding the 
propagation of CRs.

We further find that for the reacceleration model, no high-rigidity
($O(10^2)$ GV) hardening of the diffusion coefficient is necessary to 
account for the spectral hardenings of both the primary and secondary 
nuclei. The observed spectral hardenings can be largely due to the 
hardening of the injection spectrum. The reacceleration effect
automatially explain that the secondary CR spectra harden more than 
the primary ones through a steepening of the low-energy part of
the spectra. For the convection model, the inclusion of a spectral 
break of the diffusion coefficient around hundreds of GV rigidities
can improve the fit moderately. Compared with the reacceleration model, 
the convection model still fits the data much poorer even with such a 
break of the rigidity dependence of the diffusion coefficient.

Antiprotons can also be used as a diagnostics of the propagation model.
We have done fittings to the proton spectra measured by Voyager-1 
\cite{2016ApJ...831...18C}, AMS-02 \cite{2015PhRvL.114q1103A}, and
DAMPE \cite{2019SciA....5.3793A}, and the antiproton spectrum by
AMS-02 \cite{2016PhRvL.117i1103A} with the convection and reacceleration 
models based on the mean propagation parameters given in Table 
\ref{table:para}. The proton injection spectrum is approached with the 
similar spline interpolation method. An additional re-normalization
parameter of the calculated secondary antiproton spectrum and a 
force-field solar modulation potential are employed. The fit of the 
reacceleration model gives a minimum $\chi^2_{p+\bar{p}}=243.4$
for a dof of 151, which is better than that of the convection 
model (with a minimum $\chi^2_{p+\bar{p}}=316.1$). Both fits 
are not good enough. Note, however, there are complications from 
the uncertainties of the propagation parameters, the antiproton 
production cross section, the charge-sign dependence of the solar 
modulation, and/or possible exotic contribution from e.g., 
the dark matter. More careful dedicated studies will be helpful.

Although there are quite a number of discussions in literature to 
extend the propagation of CRs with more complicated configurations,
such as the spatial variation of the propagation properties
\cite{2012ApJ...752L..13T,2016ApJ...819...54G} and the anisotropic 
diffusion with respect to ordered magnetic fields 
\cite{2017JCAP...10..019C}, our results show that a simple 
two-dimensional, isotropic, and uniform propagation scenario can
give quite good description to the locally measured CRs.
Better understanding of the CR transport may be achieved with more
precise measurements of the CR distribution in the Galaxy, by e.g.,
$\gamma$-rays.

\section*{Acknowledgments}
We thank V. Ptuskin for helpful discussion.
This work is supported by the National Key Research and Development Program 
of China (No. 2016YFA0400200), the National Natural Science Foundation of 
China (Nos. 11722328, 11851305, U1738205, U1738209), the Key Research Program 
of Frontier Sciences of Chinese Academy of Sciences (No. QYZDJ-SSW-SYS024).
QY is also supported by the 100 Talents program of Chinese Academy of 
Sciences and the Program for Innovative Talents and Entrepreneur in Jiangsu.


\providecommand{\href}[2]{#2}\begingroup\raggedright\endgroup

\end{document}